\documentclass[twocolumn,
    showpacs,
    preprintnumbers,
    amsmath,
    amssymb,
    nofootinbib,
    tightenlines,
    nobibnotes,
    floats,
    prb]{revtex4-1}
\usepackage{amsfonts}
\usepackage{amsmath}
\usepackage[english]{babel}
\usepackage[T1]{fontenc}
\usepackage{times}
\usepackage{mathrsfs}
\usepackage{braket}
\usepackage{graphicx}
\usepackage{dcolumn}% Align table columns on decimal point
\usepackage{bm}% bold math
\usepackage[tight, FIGTOPCAP, hang, raggedright, nooneline]{subfigure}
\usepackage[colorlinks=true,citecolor=blue,linkcolor=blue,urlcolor=blue]{hyperref}
\usepackage{blindtext}
\usepackage{soul}
\usepackage{color}
\usepackage[dvipsnames]{xcolor}

\newcommand{\vS}{\vec S}
\newcommand{\ve}{\vec e}
\newcommand{\vj}{\vec j}
\newcommand{\up}{\uparrow}
\newcommand{\down}{\downarrow}
\newcommand{\vsigma}{\mbox{\boldmath $\sigma$}}
\newcommand{\jj}{I}
\newcommand{\II}{{\cal I}}

\usepackage{color}
\renewcommand{\vec}[1]{\boldsymbol{\mathbf{#1}}}

\begin{document}

\title{Current-induced switching of magnetic molecules on topological insulator surfaces}
\author{Elina Locane and Piet W. Brouwer}
\affiliation{Dahlem Center for Complex Quantum Systems and Institut f\"ur Theoretische Physik, Freie Universit\"at Berlin, Arnimallee 14, 14195 Berlin, Germany}
\date{\today}

\begin{abstract}

Electrical currents at the surface or edge of a topological insulator are intrinsically spin-polarized. We show that such surface/edge currents can be used to switch the orientation of a molecular magnet weakly coupled to the surface or edge of a topological insulator. For the edge of a two-dimensional topological insulator as well as for the surface of a three-dimensional topological insulator the application of a well-chosen surface/edge current can lead to a complete polarization of the molecule if the molecule's magnetic anisotropy axis is appropriately aligned with the current direction. For a generic orientation of the molecule a nonzero but incomplete polarization is obtained. We calculate the probability distribution of the magnetic states and the switching rates as a function of the applied current.
\end{abstract}

%\pacs{...}
\maketitle

\section{Introduction}

The central idea behind the field of spintronics is to use the electron's spin degree of freedom, not its charge, for information storage and processing.\cite{review_2004,annureview_2010,review_2015} Since the energy required to generate magnetic fields scales unfavorably at small length scales, electrical mechanisms for the manipulation and detection of magnetic moments are crucial for successful spintronics applications. The spin transfer torque\cite{slonczewski_1996,berger_1996} has been established as a reliable effect to manipulate the magnetization of a thin ferromagnetic layer with a spin-polarized current.\cite{ralph_stiles_2008} Whereas injection from a ferromagnet was used as a spin-polarized current source in the original realization, devices based on spin-orbit coupling have also been realized.\cite{ando_2008,pi_2010,liu_2011,miron_2011,liu_2012,kim_2013,fan_2013,garello_2013,emori_2013,ryu_2013}

An extreme form of spin-orbit coupling exists in the surface states of topological insulators (TIs):\cite{TI_fu_kane_mele,TI_hasan_kane} These states have complete spin-momentum locking, {\em i.e.}, the electron spin and its direction of motion are perfectly correlated. One consequence of the spin-momentum locking is that a surface charge current in a TI is automatically spin-polarized. Recent proposals have concentrated on exploiting this effect to control the dynamics of thin magnetic layers in the proximity of the surface of a three-dimensional TI\cite{tserkovnyak_2012,chen_2014,mahfouzi_2012,mahfouzi_2016,fischer_2016,ndiaye_2015,reza_2016} or magnets coupled to the edge of a two-dimensional TI.\cite{meng_2014,silvestrov_2016} Experiments on metallic magnetic layers in contact to a TI have reported a spin-transfer torque exceeding the values found in non-topological spin-orbit materials.\cite{fan_2014,mellnik_2014,wang_2015}

In this paper, we investigate the possibility to use the spin-polarized surface currents of a topological insulator to control the magnetic moment of a molecular magnet adsorbed on the surface of a three-dimensional topological insulator or at the edge of a two-dimensional topological insulator. Like a ferromagnet, a molecular magnet has degenerate magnetic ground states, separated by an energy barrier,\cite{bogani_2008,gatteschi_2006} although in the case of a molecular magnet the barrier is microscopic, not macroscopic, which leads to a finite relaxation time of the molecule's magnetic moment. Of particular interest are the so-called ``single-molecule magnets'',\cite{friedman_2010} which consist of a magnetic core with a spin $S \sim 6 - 12$, shielded from the environment by a (typically) organic ligand. For single-molecule magnets magnetic lifetimes of several years have been reported for temperatures below the ``blocking temperature'' set by the anisotropy barrier between the two magnetic ground states.\cite{christou_2000} Single-molecule magnets have been shown to preserve their magnetic properties, including long magnetic lifetimes, when adsorbed on conducting surfaces.\cite{candini_2011,mannini_2009,kahle_2012,malavolti_2015,margheriti_2009}

\begin{figure}[b]
    \begin{center}
        \includegraphics[width=0.45\textwidth]{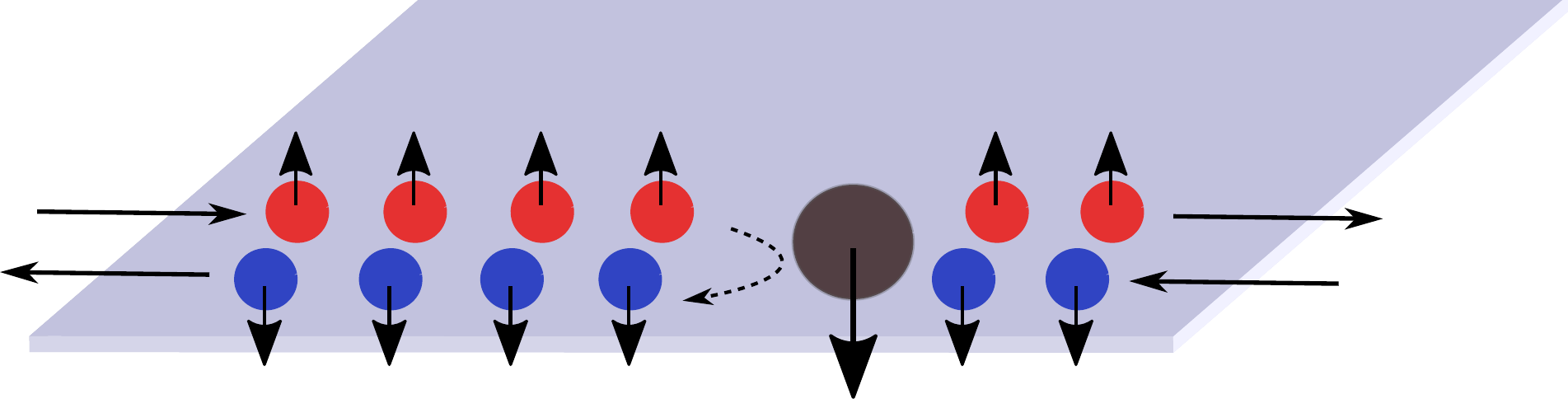}
    \end{center}
    \caption{(Color online) Current-induced switching of a localized spin $1/2$ weakly coupled to the helical edge of a two-dimensional topological insulator. Backscattering of a right-moving electron is accompanied by a flip to the spin ``up'' state of the localized spin, leaving it in a fully polarized state after a single backscattering event. For a sufficiently large applied current the reverse scattering process, which would return the localized spin to the ``down'' state, is strongly suppressed.}
    \label{fig:sketch}
\end{figure}

The mechanism by which an applied electrical current at the surface of a TI can be used to switch the orientation of a molecular magnet is best illustrated using the example of a localized spin $1/2$ coupled to the edge of a two-dimenstional topological insulator.\cite{maciejko_2009,tanaka_2011,probst_2015} The electronic state at the edge of a two-dimensional TI is helical, so that electrons moving in opposite directions have opposite spin.\cite{TI_fu_kane_mele,TI_hasan_kane} Hence, a particle current to the right is polarized as ``spin up'', see Fig.\ \ref{fig:sketch}. By spin conservation, backscattering of a right-moving electron into a left-moving state requires the localized spin to flip from down to up, so that a single backscattering event is sufficient to polarize the spin $1/2$ in the ``up'' state. If the applied current is sufficiently large, the Pauli principle forbids backscattering of left-moving electrons --- which would be accompanied by a spin flip in the opposite direction ---, so that the spin $1/2$ remains in the ``up'' state as long as the current is being applied.

This simple picture needs to be refined for molecules with a higher spin and for molecular magnets adsorbed to the two-dimensional surface of a three-dimensional TI. Unlike the spin $1/2$ of the example of Fig.\ \ref{fig:sketch}, a higher-spin molecule comes with its own anisotropy axis, and the argument based on spin conservation no longer applies in this simple form if the molecule's anisotropy axis is not aligned with the spin quantization axis of the TI edge state. Simple spin conservation arguments cannot be applied to a molecular magnet on the surface of a three-dimensional topological insulator either, because in that case there is no unique spin quantization axis, as electrons can propagate in an arbitrary direction along the surface. As we will show below, in this generic situation current-induced switching of molecular magnet on a TI surface is not perfect, although appreciable polarizations can be achieved even for randomly oriented molecules. An interesting observation is that, unlike in the example above, for a molecular magnet on the surface of a three-dimensional TI the current-induced polarization is not a monotonously increasing function of the applied current, but has a maximum at intermediate current densities.

This article is organized as follows: In Sec.\ \ref{sec:2dTI} we consider a molecular magnet at the edge of a two-dimensional TI, expanding the example of Fig.\ \ref{fig:sketch} to higher-spin molecules. We calculate the probability with which current-induced switching takes place, the current-induced switching rate, and the zero-current relaxation rate that arises from the exchange coupling of the molecule's spin to the TI edge. In Sec.\ \ref{sec:3dTI} we consider the same questions for a molecular magnet on the surface of a three-dimensional topological insulator, which is the more likely candidate for an experimental realization. We conclude with a brief outlook in Section \ref{sec:Conclusion}. The main text mainly focuses on molecules with integer spin $S$. The case of half-integer spin, which requires a technically more demanding analysis, is discussed in the Appendix.

\section{Molecular magnet at edge of a two-dimensional TI}
\label{sec:2dTI}

\subsection{Spin $1/2$}

To introduce our notation and to provide a reference for further calculations, we start by considering a localized moment of spin $S=1/2$ exchange-coupled to the helical edge of a two-dimensional topological insulator. We assume that there is only a single localized spin coupled to the edge and that the temperature is large enough that the Kondo effect can be neglected. In that case transitions between different spin states can be described using rate equations.

The coupling between the edge and the magnetic moment is described by the exchange Hamiltonian 
\begin{equation}
  H_{\rm ex} = v J \delta(z) \vec{S} \cdot \vsigma,
  \label{eq:Hex2d}
\end{equation}
where $z$ labels the coordinate along the helical edge, $v$ is the velocity of the helical edge state, and $J$ is the dimensionless exchange coupling. Because of spin-momentum locking of the edge states, backscattering of edge electrons from the molecular magnet involves simultaneous flips of the spin of the edge electrons and of the localized spin of the molecular magnet. We fix the spin $z$ axis along the direction of the spin quantization axis for the edge-state electrons,
\begin{equation}
  H_{\rm edge} = v p_z \sigma_z,
\end{equation}
so that right-moving electrons at the edge have spin up and left-moving electrons have spin down. We note that the spin $z$ axis need not be in the same direction as the laboratory $z$ axis. (In fact, for the edge of a two-dimensional topological insulator, the spin quantization axis is commonly taken to be perpendicular to the plane of the TI.\cite{bruene_2012})
The states $\ket{k\pm}$ in the helical edge are labeled by their energy $\varepsilon = \hbar v k$ and the propagation direction $\tau = \pm$, where we take $\tau = +$ for right-moving electrons (positive $z$ direction) and $\tau = -$ for left-moving electrons. (Note that in this notation $k$ merely parameterizes the energy; its magnitude $|\hbar k|$ equals the magnitude of the momentum, but the sign of $k$ is that of the energy $\varepsilon$, not of the momentum.)

The transition rates $\Gamma_{s,s'}$ between the spin states are calculated from Fermi's golden rule. The rate $\Gamma_{\downarrow,\uparrow}$ for transitions from the spin up state $\ket{\uparrow}$ to the spin down state $\ket{\downarrow}$ is
\begin{eqnarray}
  \Gamma_{\downarrow,\uparrow} &=& 
  \frac{2 \pi}{\hbar} 
  \int\limits_{-\infty}^{+\infty}\frac{dk dk'}{(2 \pi)^2}
  n_{-}(\varepsilon_k) [1 - n_{+}(\varepsilon_{k'})]
  \nonumber \\ && \mbox{} \times
  |\langle +,\downarrow|H_{\rm ex}| -,\uparrow \rangle|^2 
  \delta(\varepsilon_k - \varepsilon_{k'})
  \nonumber \\ &=&
  \frac{J^2}{2 \pi \hbar} \int\limits_{-\infty}^{+\infty} d \varepsilon
  n_{-}(\varepsilon) (1 - n_{+}(\varepsilon)).
  \label{eq:Gamma1}
\end{eqnarray}
(The matrix elements of $H_{\rm ex}$ do not depend on $k$ and $k'$, which is why we have suppressed $k$, $k'$ in our notation.)
Similarly, the rate $\Gamma_{\uparrow,\downarrow}$ at which transitions from the spin down state to the spin up state take place reads
\begin{equation}
  \Gamma_{\uparrow,\downarrow} = 
  \frac{J^2}{2 \pi \hbar} \int\limits_{-\infty}^{+\infty} d \varepsilon
  n_{+}(\varepsilon) (1 - n_{-}(\varepsilon)).
  \label{eq:Gamma2}
\end{equation}

The distribution function at temperature $T$ and chemical potential $\mu$ in the presence of an edge current $\II = \jj e$ is
\begin{eqnarray}
  n_{\tau}(\varepsilon) &=& 
  \frac{1}{1 + e^{(\varepsilon - \mu - \tau \pi \hbar \jj)/k_{\rm B} T}},
\end{eqnarray}
as one can easily verify from the relation $\jj = (1/2 \pi \hbar) \int d\varepsilon [n_{+}(\varepsilon) - n_{-}(\varepsilon)]$. For definiteness, we will assume throughout that the particle current is to the right, $\jj > 0$. Performing the integrations in Eqs.\ (\ref{eq:Gamma1}) and (\ref{eq:Gamma2}) then gives
\begin{equation}
  \Gamma_{\downarrow,\uparrow} =
  J^2 g(\jj),\ \
  \Gamma_{\uparrow,\downarrow} = J^2 g(-\jj),
  \label{eq:Gamma12}
\end{equation}
where we abbreviated
\begin{equation}
  g(\jj) = \frac{\jj}{e^{2 \pi \hbar \jj/k_{\rm B} T} - 1}.
  \label{eq:g}
\end{equation}
For high applied currents, $\hbar \jj \gg k_{\rm B} T$, scattering processes in which a left-moving electron is scattered into a right-moving one and, correspondingly, the rate $\Gamma_{\downarrow,\uparrow}$ are exponentially suppressed. On the other hand, for low applied currents $|\jj| \ll k_{\rm B} T /\hbar$, scattering of thermally excited charge carriers dominates the spin switching rates, and the application of the current $\jj$ only gives a slight asymmetry.

In the absence of coherences between the spin states $\ket{1/2}$ and $\ket{-1/2}$ the probabilities $P_{\uparrow}$ and $P_{\downarrow}$ to find the spin a state of spin $s = \uparrow, \downarrow$ can be solved from the stationary solutions of the rate equation
\begin{align}
  \frac{d P_{s}}{dt} = \sum_{s'\neq s}
  ( \Gamma_{s,s'} P_{s'} - \Gamma_{s',s} P_s) .
  \label{eq:master_time-dep}
\end{align}
(In principle, one needs a density matrix to describe possible coherent superpositions of the degenerate states $\ket{1/2}$ and $\ket{-1/2}$, see, {\em e.g.}, Refs.\ \onlinecite{nazarov_1993,gurvitz_1998,koenig_2003,braun_2004,braig_brouwer_2005,probst_2015} and the Appendix. However, such coherences do not occur in the present case, in which there is no external magnetic field and a single spin quantization axis for the conduction electrons at the TI edge and the localized spin.) Substituting the rates (\ref{eq:Gamma12}) one immediately finds
\begin{equation}
  P_{\uparrow} = 1 - P_{\downarrow} =
  \frac{1}{1 + e^{-2 \pi \hbar \jj/k_{\rm B} T}}.
\end{equation}
The same result also follows from the observation that the coupling to the conduction electrons at the TI edge effectively amounts to a Zeeman shift $2 \pi \hbar \jj$ between the spin up and spin down states of the molecule. We conclude that the application of a current $\jj \gg k_{\rm B} T/\hbar$ leads to a complete polarization of the molecule's spin.

\subsection{Higher-spin molecule}

Although the application of a current through the helical edge of a two-dimensional topological insulator causes a polarization of a spin $1/2$ exchange-coupled to the edge state, the induced magnetic moment quickly disappears as soon as the current is switched off. In our model, in which the coupling to the helical edge state is the only source of relaxation, the corresponding relaxation rate $\Gamma$ is given by the low-current limit of Eq.\ (\ref{eq:Gamma12}),
\begin{equation}
  \Gamma_{\uparrow,\downarrow} = \Gamma_{\downarrow,\uparrow} =
  \frac{J^2 k_{\rm B} T}{2 \pi \hbar}.
\end{equation}

\begin{figure}
     \includegraphics[width=0.45\textwidth]{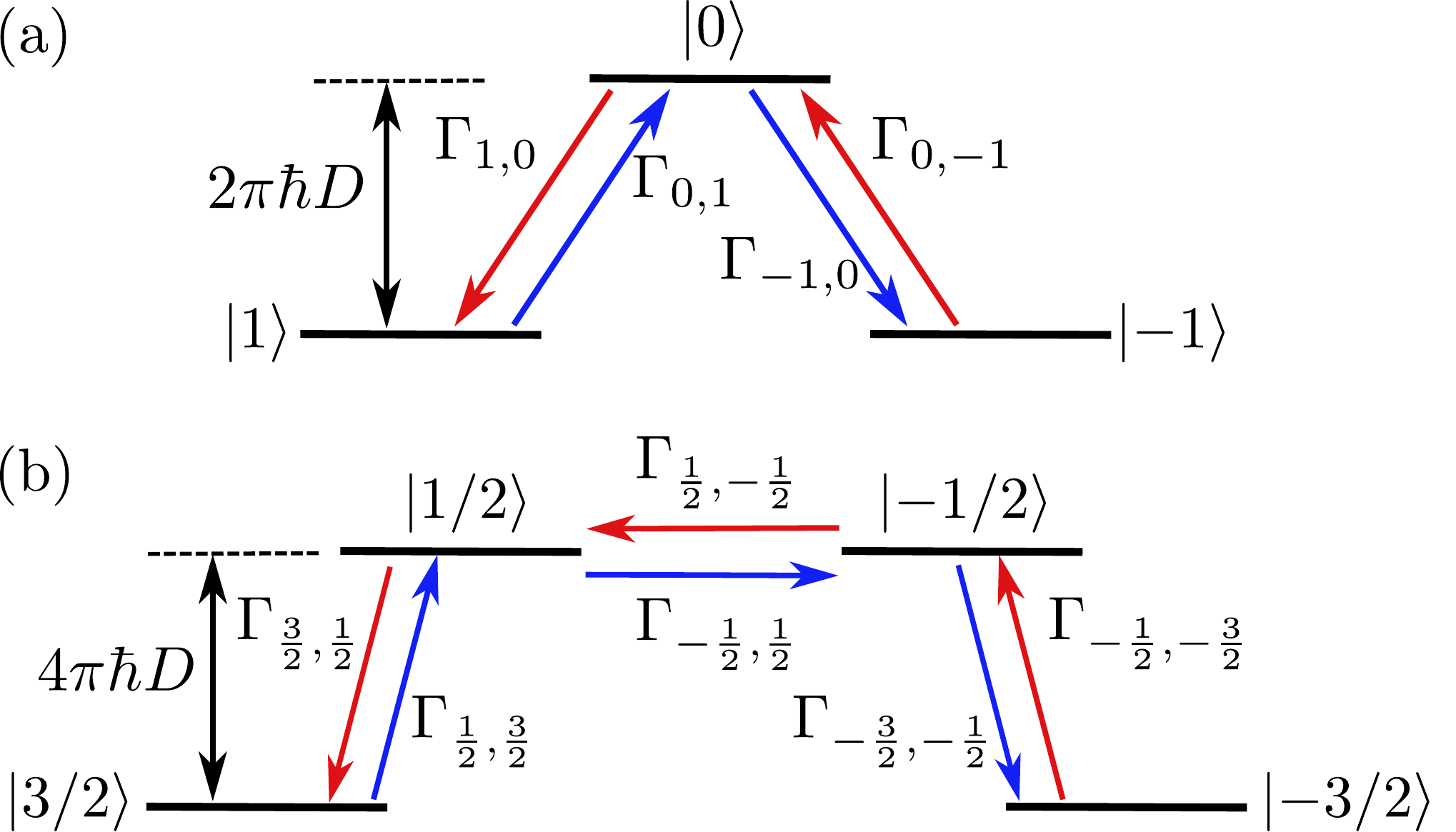}
    \caption{(Color online) Schematic drawing of energy levels and transition rates $\Gamma_{s \pm 1, s}$ (a) for a spin-$1$ molecule and (b) for a spin-$3/2$ molecule.}
    \label{fig:sketch_spin1}
\end{figure}

Longer relaxation times in the absence of an applied current requires molecules with a higher spin $S$. In this case, magnetic anisotropy,
\begin{equation}
  H_{\rm anis} = - \frac{2 \pi}{\hbar} D (\vS \cdot \ve)^2,
  \label{eq:Hs}
\end{equation}
imposes an energy barrier between states with maximal and minimal spin $\pm \hbar S$ (measured along the anisotropy axis $\ve$), see Fig.\ \ref{fig:sketch_spin1}, which, if $k_{\rm B} T$ is lower than the barrier energy, leads to strongly enhanced lifetimes. (See below for a quantitative estimate for our model.) In our analysis, we have adopted a simple easy-axis anisotropy strength $D$ (which we take to have the units of frequency) with $D > 0$, although our considerations also carry over to other forms for the anisotropy energy in which there are two different minima. (The case $D < 0$ is for easy-plane anisotropy, which does not have two separated energy minima and will not be considered here.) We label the spin states with the quantum number $s = -S$, $-S+1$, \ldots, $S$ for the spin component $\hbar s$ along the $\ve$-direction. (For $S=1/2$, $s$ takes the values $s=\pm 1/2$, corresponding to the notation ``$\up$'' and ``$\down$'' in the previous Subsection.)

If the anisotropy axis is aligned with the spin quantization axis for the helical edge state, the analysis of the previous Subsection immediately carries over, and one finds
\begin{equation}
%  \Gamma_{s - 1,s} &=& 
%  J^2 (S+s)(S-s+1)
%  g[j + D (2 s - 1)], \nonumber \\
  \Gamma_{s \pm 1,s} =
  J^2 (S \mp s)(S \pm s+1)
  g[\mp(\jj + D (2 s \pm 1)].
  \label{eq:Gammass}
\end{equation}
For the probabilities $P_s$ one then finds
\begin{equation}
  P_{s} = \frac{1}{Z} 
  e^{(2 \pi s \hbar \jj + 2 \pi \hbar D s^2)/k_{\rm B} T},
\end{equation}
where the prefactor is fixed by the normalization condition $\sum_{s} P_{s} = 1$. 

If the temperature is much smaller than the anisotropy energy, which is a condition that we will assume throughout this article, the rate equations can be solved directly, using detailed balance, to give the zero-current switching rate $\Gamma_{\rm switch}(0)$ between the two magnetic ground states $\ket{S}$ and $\ket{-S}$,
\begin{equation}
  \Gamma_{\rm switch}(0) \approx \frac{1}{2} J^2 S(S+1) D e^{-2 \pi \hbar D S^2/k_{\rm B} T} \label{eq:GammaSwitch0Even}
\end{equation}
if $S$ is integer and 
\begin{eqnarray}
  \Gamma_{\rm switch}(0) & \approx & J^2 (S+1/2)^2 
  \frac{k_{\rm B} T}{2 \pi \hbar} 
  \nonumber \\ && \mbox{} \times 
  e^{- 2 \pi \hbar D (S^2-1/4)/k_{\rm B} T}
  \label{eq:GammaSwitch0Odd}
\end{eqnarray}
if $S$ is half integer. 

The application of a current $\jj$ slightly larger than $D (2 S - 1)$ leads to a quick and complete spin polarization of the molecule. The switching rate $\Gamma_{\rm switch}(\jj)$ can be estimated as the inverse of the sum of inverse transition rates $\Gamma_{s,s+1}$, which gives 
\begin{equation}
  \Gamma_{\rm switch}(\jj) \sim J^2 \jj (S+1/2)/ \ln (4 S + 1)
  \label{eq:GammaSwitch2dHigh}
\end{equation}
for large $S$. A smaller current $k_{\rm B} T \ll \hbar \jj \ll \hbar D (2 S + 1)$ also polarizes the molecule, but since thermal activation is still needed in the switching process, the time required to reach the polarized state is still large (though much shorter than the zero-current lifetime),
\begin{equation}
  \Gamma_{\rm switch}(\jj) \approx 
  \Gamma_{\rm switch}(0) e^{2 \pi \hbar \jj S/k_{\rm B} T}
%  \frac{1}{2} J^2 S(S+1) D e^{2 \pi \hbar(\jj S - D S^2)/k_{\rm B} T}
\end{equation}
if $S$ is integer and 
\begin{eqnarray}
  \Gamma_{\rm switch}(\jj) & \approx & 
  \Gamma_{\rm switch}(0) e^{2 \pi \hbar \jj (S-1/2)/k_{\rm B} T}
\end{eqnarray}
if $S$ is half integer. 

If $\ve$ is not aligned with the $z$ axis, the expression for the rates $\Gamma_{s \pm 1,s}$ involves the matrix elements
\begin{widetext}
\begin{eqnarray}
  \langle \tau',s-1|H_{\rm ex}|\tau , s \rangle &=&
  \frac{J v \hbar}{2} \sqrt{(S+s)(S-s+1)}
%  \times 
  \left\{ \begin{array}{ll}
  - \tau \sin \theta & \mbox{if $\tau' = \tau$}, \\
  \tau' (1 + \tau' \cos \theta) e^{-i \phi \tau'}
  & \mbox{if $\tau' = - \tau$},
  \end{array} \right. \nonumber \\
  \langle \tau',s+1|H_{\rm ex}|\tau, s \rangle &=&
  \frac{J v \hbar}{2} \sqrt{(S-s)(S+s+1)}
%  \times 
  \left\{ \begin{array}{ll}
  - \tau \sin \theta & \mbox{if $\tau' = \tau$}, \\
  \tau (1 + \tau \cos \theta) e^{i \phi \tau} & \mbox{if $\tau' = - \tau$},
  \end{array} \right.
  \label{eq:MatrixElements2d}
\end{eqnarray}
where $\theta$ and $\phi$ are the polar angles corresponding to the anisotropy axis $\ve = \cos \theta \ve_z + \sin \theta \cos \phi \ve_x + \sin \theta \sin \phi \ve_y$. This gives
\begin{eqnarray}
  \Gamma_{s \pm 1,s} &=&
  \frac{1}{4} J^2 (S \mp s) (S \pm s+1) 
  \sum_{\tau}
  \left\{
   g[\mp D (2 s \pm 1)] \sin^2 \theta
  + (1 \mp \tau \cos \theta)^2 
  g[\tau \jj \mp D (2 s \pm 1)]
  \right\}. \label{eq:GammaAnis}
\end{eqnarray}
One verifies that the zero-current rates are the same as those of Eqs.\ (\ref{eq:Gammass}), independent of the direction $\ve$ of the anisotropy axis, so that Eqs.\ (\ref{eq:GammaSwitch0Even}) and (\ref{eq:GammaSwitch0Odd}) still apply. 

Because of the degeneracy of the states $\ket{1/2}$ and $\ket{-1/2}$ the rate equation (\ref{eq:master_time-dep}) is not sufficient to describe transitions between these two states and, instead, one has to use a master equation for the reduced $2 \times 2$ density matrix that is able to account for coherent superpositions of the states $\ket{1/2}$ and $\ket{-1/2}$. For that reason we here restrict ourselves to the case of integer $S$, for which this complication does not occur. A complete analysis and final expressions for half-integer $S$ are given in the Appendix.

We first discuss the case of an ``intermediate'' current $k_{\rm B} T \ll \hbar \jj \ll \hbar D (2 S - 1)$, for which the transition rates (\ref{eq:GammaAnis}) can be approximated as
\begin{eqnarray}
  \Gamma_{s+1,s} & \approx &
  \frac{ J^2 D}{4} (S-s)(S+s+1) |2 s + 1|
  \left\{ \begin{array}{ll} 4 & \mbox{if $s \ge 0$}, \\ 
  (1 + \cos \theta)^2 
    e^{2 \pi \hbar (\jj - D|2 s + 1|)/k_{\rm B} T}
  & \mbox{if $s < 0$},
  \end{array} \right.
  \nonumber \\ 
  \Gamma_{s-1,s} & \approx &
  \frac{J^2 D}{4} (S+s)(S-s+1) |2 s - 1|
  \left\{ \begin{array}{ll} 
  (1 - \cos \theta)^2
    e^{2 \pi \hbar (\jj - D|2 s - 1|)/k_{\rm B} T}
  & \mbox{if $s > 0$}, \\ 4 & \mbox{if $s \le 0$}. 
  \end{array} \right.
\end{eqnarray}
\end{widetext}
The large difference between ``upstream'' transitions to higher-energy spin states and ``downstream'' transitions to lower-energy spin states ensures that the molecule will predominantly be in one of its two magnetic ground states $\ket{S}$ and $\ket{-S}$. The presence of the applied current breaks the symmetry between these two ground states and leads to a preferred population of the state $\ket{S}$ if $\theta < \pi/2$ and $\ket{-S}$ if $\theta > \pi/2$, although there is no longer a perfect polarization for generic $\theta$,
\begin{eqnarray}
   P_{S} &=& 1 - P_{-S} \nonumber \\ &=&
   \frac{(1 + \cos \theta)^{2 S}}{(1 + \cos \theta)^{2 S}
  + (1 - \cos \theta)^{2 S}}.
  \label{eq:P2dIntermediateEven}
\end{eqnarray}
Switching between the magnetic ground states involves thermal activation, which is why the switching rate between the magnetic ground states is exponentially long in $2 \pi \hbar D S^2/k_{\rm B} T$, although the rate is parametricaly larger than the spontaneous relaxation rate at zero current. Defining $\Gamma_{\rm switch}(\jj)$ as the switching rate from $\ket{-S}$ to $\ket{S}$ we have
\begin{equation}
  \Gamma_{\rm switch}(\jj) =
  \Gamma_{\rm switch}(0) \left( \frac{1 + \cos \theta}{2} \right)^{2S}
  e^{2 \pi \hbar \jj S/k_{\rm B} T},
\end{equation}
where $\Gamma_{\rm switch}(0)$ is the zero-current relaxation rate of Eq.\ (\ref{eq:GammaSwitch0Even}). The switching rate for the inverse process is obtained upon replacing $\theta \to \pi - \theta$.
\begin{figure}
   \includegraphics[width=0.45\textwidth]{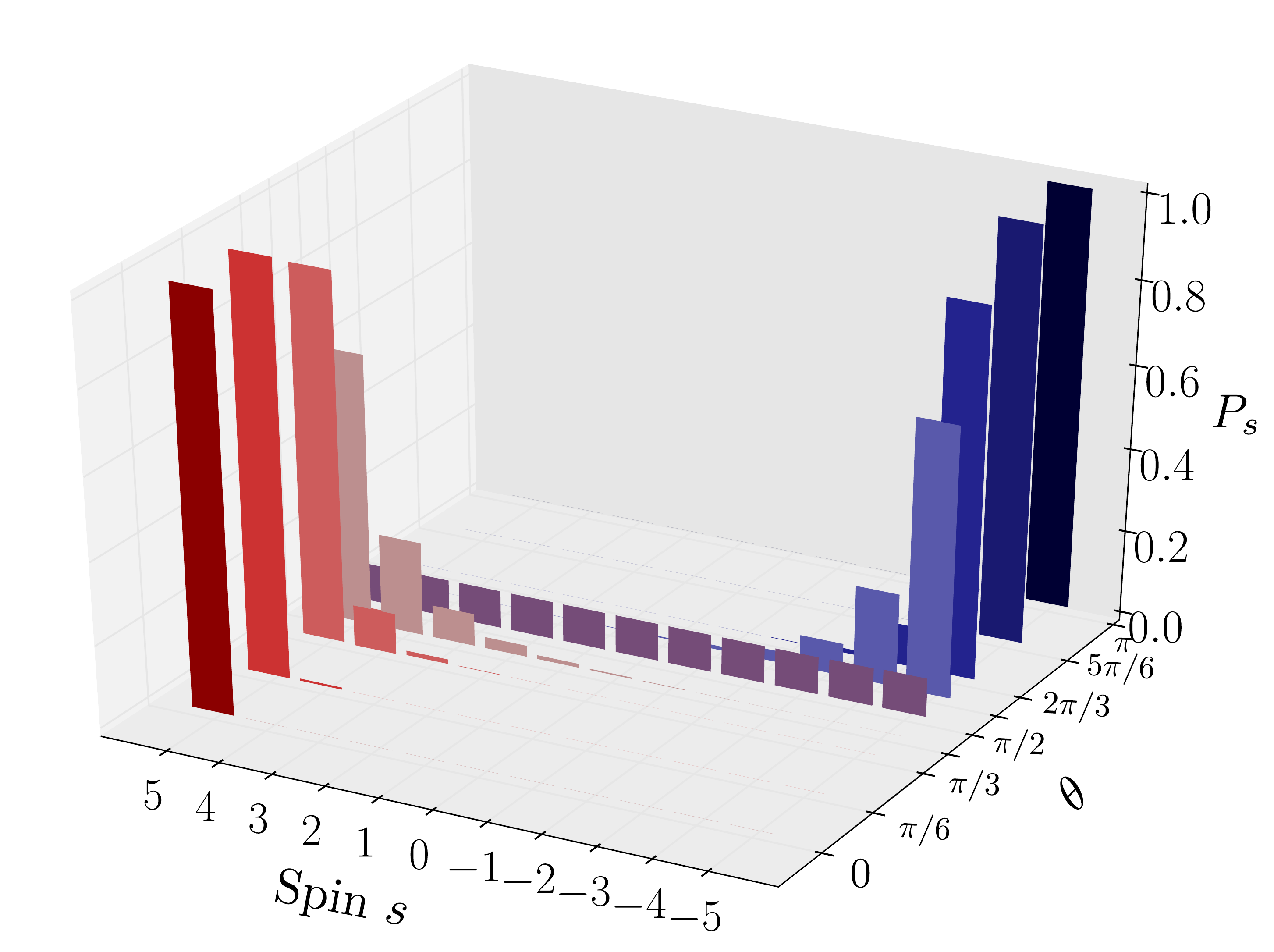}
    \caption{(Color online) Steady-state probability distribution (\ref{eq:Ps2d}) for $S=5$ and for angles $\theta=0$, $\theta = \pi/6$, $\theta = \pi/3$, $\theta = 5\pi/12$, $\theta = \pi/2$, $\theta = 7\pi/12$, $\theta = 2\pi/3$, $\theta = 5\pi/6$ and $\theta = \pi$.}
    \label{fig:Ps2d}
\end{figure}
The difference between the case $\theta=0$ and generic $\theta$ is most pronounced in the ``high-current'' limit $\jj \gg D (2 S - 1)$, in which the rates $\Gamma_{s \pm 1,s}$ in Eq.\ (\ref{eq:GammaAnis}) are dominated by the terms proportional to $g[-\jj \mp D (2 s \pm 1)]$. Approximating $g[-\jj \pm D (2 s \mp 1)] \approx \jj$ for $\jj \gg D(2 S-1)$, we find
\begin{eqnarray}
  \Gamma_{s \pm 1,s} &=& \frac{\jj}{4} J^2 (S \mp s) (S \pm s+1) 
  (1 \pm \cos \theta)^2.
  \label{eq:GammaHigh2d}
\end{eqnarray}
For the probabilities $P_s$ this gives immediately
\begin{equation}
  P_s = \frac{4 \cos \theta (1 - \cos \theta)^{2(S - s)} (1 + \cos \theta)^{2(S + s)}}{(1 + \cos \theta)^{2(2 S + 1)} - (1 - \cos \theta)^{2(2 S + 1)}}.
  \label{eq:Ps2d}
\end{equation}
For $\theta$ close to zero this still gives a narrow distribution around $s = S$, but perfect polarization is not reached if $\theta \neq 0$ independent of the magnitude of the current $\jj$. Examples of the distribution (\ref{eq:Ps2d}) are shown in Fig.\ \ref{fig:Ps2d} for $S=5$ and for several representative angles $\theta$. 

\begin{figure}[t]
  \includegraphics[width=0.45\textwidth]{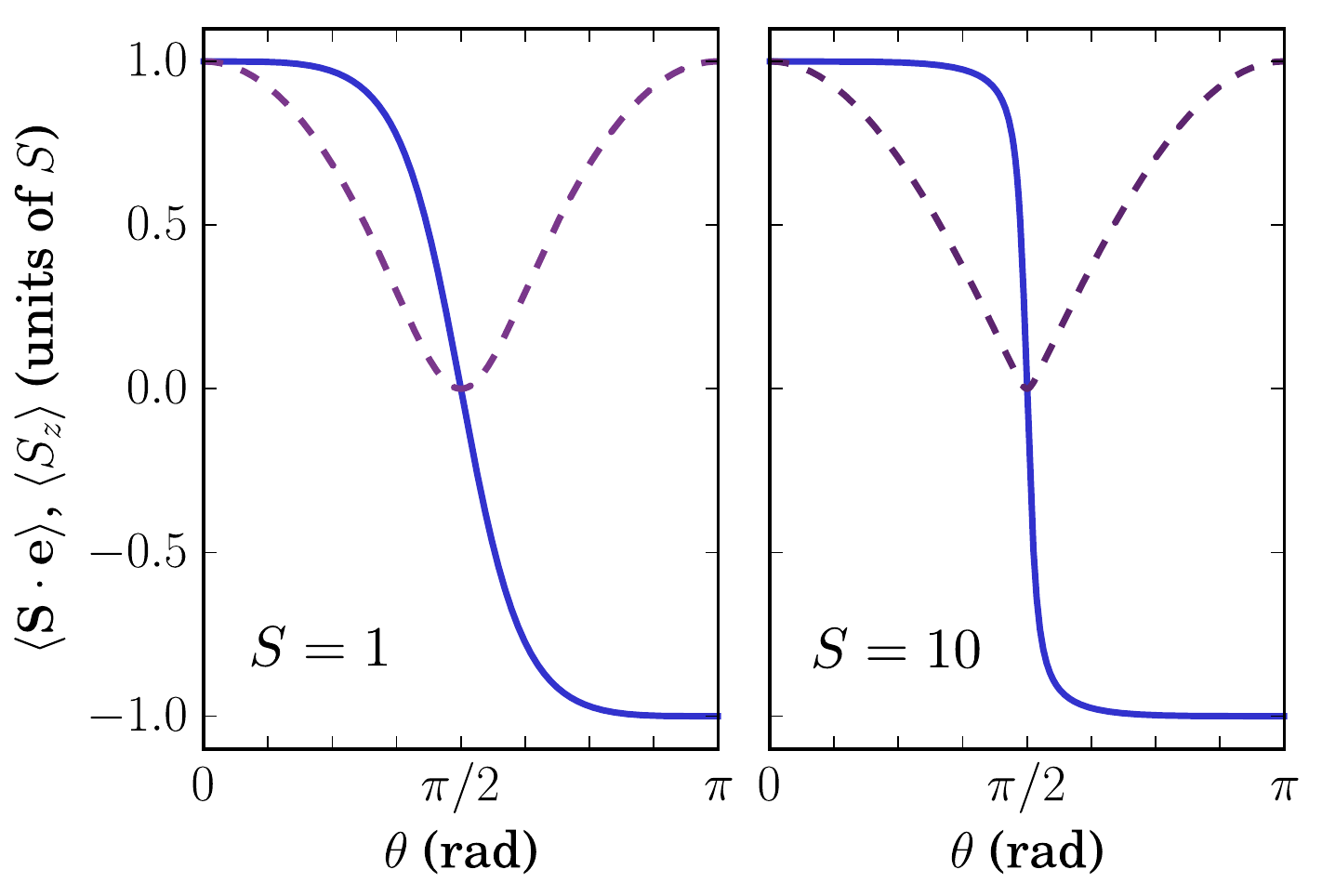}
    \caption{(Color online) Mean spin component $\langle \vS \cdot \ve \rangle$ along the direction of the anisotropy axis (solid curves) and the mean spin component $\langle S_z \rangle$ in the current direction (dashed), as a function of the angle $\theta$, for $S=1$ (left panel) and $S=10$ (right panel).}
  \label{fig:Mean2d}
\end{figure}

Figure \ref{fig:Mean2d} shows the mean values $\langle \vS \cdot \ve \rangle $ of the spin component along the anisotropy axis and $\langle S_z \rangle = \langle \vS \cdot \ve \rangle (\ve\cdot\ve_z)$ of the spin component along the current direction as a function of the angle $\theta$ for two different values of the total spin $S$. Except for $\theta = \pi/2$, the application of a current always results in a net spin in the current direction. For a randomly oriented molecule, the average moment in the current direction approaches $S/2$ in the limit of large $S$.
Without a perfect current-induced polarization, we should define the current-induced ``switching rate'' $\Gamma_{\rm switch}(\jj)$ as the rate at which the steady-state probability $P_s$ is approached, starting from $\ket{-S}$ (if $\theta < \pi/2$). The estimate (\ref{eq:GammaSwitch2dHigh}) for the switching rate $\Gamma_{\rm switch}(\jj)$ obtained above for $\ve$ parallel to the $z$ axis remains valid as a good order-of-magnitude estimate for the case of a general orientation of the anisotropy axis in the high-current regime.

\section{Three-dimensional topological insulator}\label{sec:3dTI}
The main result of the previous Section is that the current-induced polarization of a molecular magnet coupled to the edge of a two-dimensional TI is complete only if the molecule's anisotropy axis is aligned with the spin quantization axis for the TI edge state, whereas the polarization is incomplete --- but generally nonzero --- for arbitrary orientations of the molecule. The analysis was simplified by the fact that the edge of a two-dimensional TI has a unique propagation direction and, hence, a unique spin quantization axis. This is the main difference with the case of a molecule coupled to the surface of a three-dimensional TI: Electrons at the surface of a three-dimensional TI can propagate in all directions in the plane of the surface, even in the presence of a large applied current. Hence, there is not a unique spin quantization axis, and there will always be surface-state electrons with a spin that is not aligned with the anisotropy axis of the molecular magnet. Nevertheless, as we show below, even for a molecule adsorbed on the surface of a three-dimensional TI the application of an electrical current can lead to a complete spin polarization, provided the molecule's anisotropy axis is aligned with the current direction and the magnitude of the applied current is appropriately chosen. As in the case of a two-dimensional TI, the spin polarization of the molecular magnet is not complete (but generally nonzero) if the molecule's magnetic anisotropy axis is not aligned with the current direction.
\begin{figure}[b]
    \begin{center}
        \includegraphics[width=0.45\textwidth]{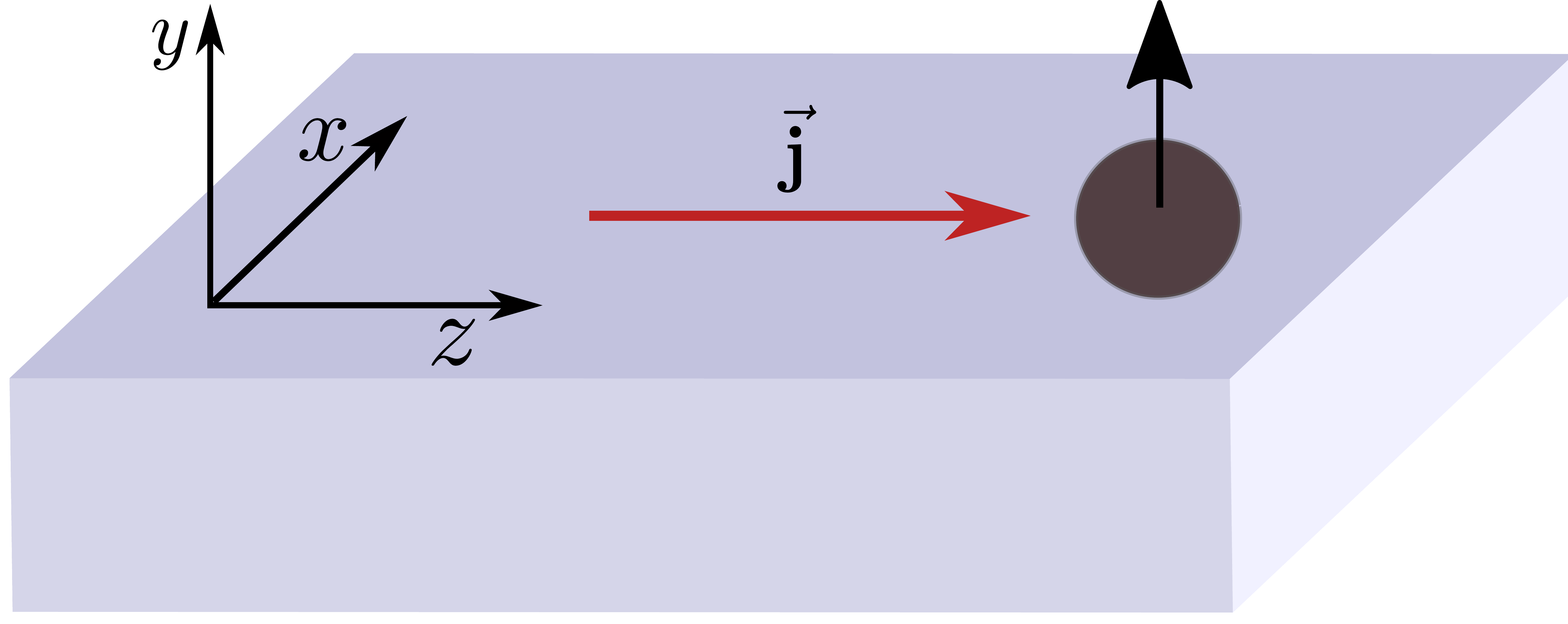}
    \end{center}
    \caption{(Color online) Molecular magnet on the surface of a three-dimensional topological insulator. The TI surface is the $xz$ plane. A current is applied the positive $z$ direction.}
    \label{fig:sketch3d}
\end{figure}
To keep the notation close to that of the previous Section, we take the TI surface to be the $xz$ plane and take the direction of (particle) current flow to be the positive $z$ direction, see Fig.\ \ref{fig:sketch3d}. The electrons at the TI surface are then described by the Hamiltonian
\begin{equation}
  H_{\rm surface} = v (p_x \sigma_x + p_z \sigma_z).
\end{equation}
We write the current density $\vj$ as
\begin{equation}
  \vj = \frac{1}{4} k_{\rm F} \jj \ve_z,
\end{equation}
where $k_{\rm F}$ is the Fermi wavenumber and $\jj$ has the dimension of current. For the exchange Hamiltonian we take
\begin{equation}
  H_{\rm ex} = \frac{2 v}{k_{\rm F}} J \delta(x) \delta(z) \vec{S} \cdot \vsigma,
  \label{eq:Hex3d}
\end{equation}
where, as in the previous Section, a prefactor has been included to make the strength $J$ of the exchange interaction dimensionless.
The Hamiltonian $H_{\rm anis}$ for the magnetic anisotropy of the magnetic molecule is the same as in the previous Subsection, see Eq.\ (\ref{eq:Hs}).
The eigenstates of $H_{\rm surface}$ are labeled by the energy $\varepsilon = \hbar v k$ and by the angle $\varphi$ of the propagation direction in the $xz$ plane, such that $\varphi = 0$ corresponds to the positive $z$ direction. We assume that the chemical potential $\mu = \hbar k_{\rm F} v \gg \max(k_{\rm B} T, \hbar \jj)$, which ensures that thermally excited carriers as well as carriers contributing to the current flow remain well away from the Dirac point. In that case the distribution function in the presence of a particle current density $1/4 k_{\rm F} \jj \ve_z$ is
\begin{equation}
  n(\varepsilon,\varphi) = \frac{1}{1 + e^{(\varepsilon - \mu - \pi \hbar \jj \cos \varphi)/k_{\rm B} T}}.
  \label{eq:n3d}
\end{equation}
Equation (\ref{eq:n3d}) both follows from a solution of the Boltzmann equation and maximizes the entropy under the constraint of a fixed current density $\vj = k_{\rm F} \jj \vec{e}_z/4$.

For the calculation of the transition rates between different spin states we need the matrix elements
\begin{widetext}
\begin{eqnarray}
  \langle \varphi',s \pm 1| H_{\rm ex} | \varphi,s \rangle &=&
  \frac{J v \hbar}{k_{\rm F}} \sqrt{(S \mp s)(S \pm s+1)} \nonumber \\ && \mbox{} \times
  \left[(\cos \theta \cos \phi \pm i \sin \phi) \sin \varphi_+
  \pm (\cos \phi \pm i \cos \theta \sin \phi) \sin \varphi_-
  - \sin \theta \cos \varphi_+ \right], 
  \label{eq:MatrixElements3d}
\end{eqnarray}
where we abbreviated $\varphi_{\pm} = (\varphi' \pm \varphi)/2$. Equation (\ref{eq:MatrixElements3d}) generalizes Eq.\ (\ref{eq:MatrixElements2d}) to the case of a surface state. As in Eq.\ (\ref{eq:MatrixElements2d}), $\theta$ and $\phi$ are the polar angles marking the direction of the anisotropy axis. For the transition rates we then find, again assuming $\mu \gg \max(k_{\rm B} T, \hbar \jj)$,
\begin{eqnarray}
  \Gamma_{s \pm 1,s} &=& J^2 (S \mp s)(S \pm s+1)
  \int \frac{d\varphi'}{2\pi} \frac{d\varphi}{2 \pi}
  (1 \pm \cos \theta \cos \varphi + \sin \theta \cos \phi \sin \varphi)
  (1 \mp \cos \theta \cos \varphi' + \sin \theta \cos \phi \sin \varphi')
  \nonumber \\ && \mbox{} \times
  g[\jj (\cos \varphi' - \cos \varphi)/2 \mp D (2 s \pm 1)],
  \label{eq:GammaGeneral}
\end{eqnarray}
where the function $g(\jj)$ was introduced in Eq.\ (\ref{eq:g}). We will now analyze these rates and the resulting probabilities $P_s$ in the regimes of low current densities, $\hbar |\jj| \ll k_{\rm B} T$, intermediate current densities, $k_{\rm B} T \ll \hbar \jj \ll \hbar D (2 S-1)$, and high current densities, $\hbar \jj \gg \hbar D (2 S-1)$. (For $S=1/2$ no intermediate regime exists and the high-current regime is defined as $\hbar \jj \gg k_{\rm B} T$.)

For low current densities $\hbar |\jj| \ll k_{\rm B} T$ the expressions (\ref{eq:GammaGeneral}) for the transition rates simplify to
\begin{eqnarray}
  \Gamma_{s+1,s} & \approx & 
  \Gamma_{-s-1,-s} \approx
  J^2 (S-s)(S+s+1) \times \left\{
  \begin{array}{ll} 
  D |2 s + 1| & \mbox{if $s > -1/2$}, \\
  k_{\rm B} T / 2 \pi \hbar & \mbox{if $s = -1/2$}, \\
  D |2 s + 1| e^{-2 \pi \hbar D |2 s + 1|/k_{\rm B} T} & \mbox{if $s < -1/2$}.
  \end{array} \right.
  \label{eq:Gamma3dLow}
\end{eqnarray}
The resulting zero-current switching rate $\Gamma_{\rm switch}(0)$ is given by the same expression as in the two-dimensional case, see Eqs.\ (\ref{eq:GammaSwitch0Even}) and (\ref{eq:GammaSwitch0Odd}).

For intermediate current densities, $k_{\rm B} T \ll \hbar \jj \ll \hbar D(2S-1)$, the exponential suppression of the transition rates to higher-energy spin states is reduced, whereas the transition rates into the lower-energy states remains approximately as in the low-current regime. For a quantitive analysis we restrict ourselves to the case of integer spin $S$, referring to the Appendix for a discussion of half-integer $S$. In the intermediate-current regime the transition rates can be approximated as 
\begin{eqnarray}
  \Gamma_{s+1,s} & \approx & 
    J^2 (S-s)(S+s+1) \times \left\{
  \begin{array}{ll} 
  D |2 s + 1| & \mbox{if $s \ge 0$}, \\
  \frac{k_{\rm B} T}{2\pi^2 \hbar \jj}
  (1 + \cos \theta)^2 D |2 s + 1|
    e^{2 \pi \hbar (\jj - D |2 s + 1|)/k_{\rm B} T} & 
  \mbox{if $s < 0$}, \end{array} \right. \nonumber \\  
  \Gamma_{s-1,s} & \approx & 
    J^2 (S+s)(S-s+1) \times \left\{
  \begin{array}{ll} 
  \frac{k_{\rm B} T}{2 \pi^2 \hbar \jj}
  (1 - \cos \theta)^2 D |2 s - 1| 
    e^{2 \pi \hbar (\jj - D |2 s - 1|)/k_{\rm B} T} & 
    \mbox{if $s > 0$}, \\
  D |2 s - 1| & \mbox{if $s \le 0$}.
 \end{array} \right. \label{eq:Gamma3dIntermediate}
\end{eqnarray}
\end{widetext}
As in the two-dimensional case, the applied current breaks the symmetry between states with positive and negative spin $s$, leading to an imbalance between the populations of the ground states $\ket{S}$ and $\ket{-S}$,
\begin{eqnarray}
   P_{S} &=& 1 - P_{-S} \nonumber \\ &=&
   \frac{(1 + \cos \theta)^{2 S}}{(1 + \cos \theta)^{2 S}
  + (1 - \cos \theta)^{2 S}}.
  \label{eq:P3dEven}
\end{eqnarray}
The population of the excited states $|s|<S$ remains exponentially small in $\hbar D (2 S - 1)/k_{\rm B} T$. The current-induced switching rate from the state $\ket{-S}$ into $\ket{S}$ is
\begin{eqnarray}
  \Gamma_{\rm switch}(\jj) &=&
  \Gamma_{\rm switch}(0) 
  \left[ \frac{k_{\rm B} T (1 + \cos \theta)^2}
  {2 \pi^2 \hbar \jj} \right]^S
  \nonumber \\ && \mbox{} \times
  e^{2 \pi \hbar S \jj/k_{\rm B} T}.
  \label{eq:GammaSwitchJ3dEven}
\end{eqnarray}
The inverse rate is given by the same expression, but with $\cos \theta$ replaced by $- \cos \theta$.

The case $\theta = 0$ that the molecule's anisotropy axis is aligned with the current direction requires to be considered separately. In this case the leading approximation of Eq.\ (\ref{eq:Gamma3dIntermediate}) vanishes for the ``upstream'' rate $\Gamma_{s-1,s}$ with $s > 0$ and subleading contributions in the small parameter $\jj/D$ must be taken into account. One finds
\begin{eqnarray}
  \Gamma_{s-1,s} & \approx & J^2 (S+s)(S-s+1) 
  \frac{D |2 s - 1| (k_{\rm B} T)^3}{8 \pi^4 \hbar^3 \jj^3}
  \nonumber \\ && \mbox{} \times
  e^{2 \pi \hbar(\jj -D |2 s - 1|)/k_{\rm B} T} 
\end{eqnarray}
if $s > 0$. Asymptotically, for $\hbar \jj \gg k_{\rm B} T$ [but still $\jj \ll D (2 S - 1)$], this leads to a complete spin polarization of the molecule, 
\begin{equation}
  P_{s} = \delta_{S,s}.
\end{equation}
[The probability $P_{-S}$ vanishes $\propto (k_{\rm B} T/\hbar \jj)^{2 S}$; all other probabilities are exponentially small in $\hbar D (2 S - 1)/k_{\rm B} T$.]
The perfect polarization can be understood if one observes that the surface-state electrons responsible for the transitions to higher-energy spin states are predominantly electrons moving in the positive $z$ direction, which are then backscattered to the negative $z$ direction. These electrons have a unique spin quantization axis, and their scattering leads to a well-defined change of the molecule's spin state $\ket{s}$. The switching rate from the state $\ket{-S}$ to $\ket{S}$ is given by Eq.\ (\ref{eq:GammaSwitchJ3dEven}) with $\theta = 0$. The rate for the inverse process $\ket{S} \to - \ket{-S}$ vanishes in the limit $k_{\rm B} T \ll \hbar \jj$.

\begin{figure}[t]
	\includegraphics[width=0.45\textwidth]{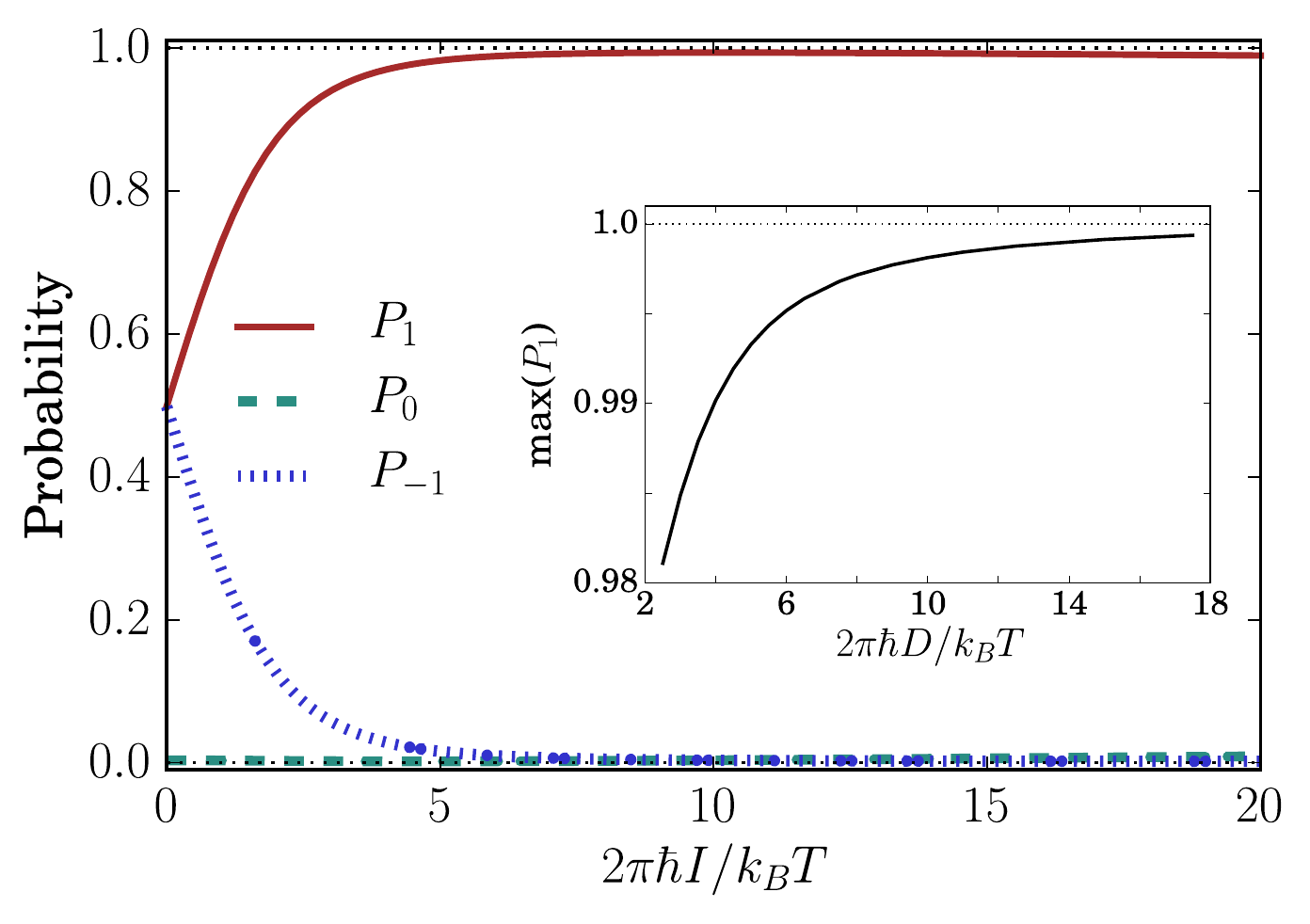}
    \caption{(Color online) The probabilities $P_{1}$, $P_{0}$, and $P_{-1}$ for a spin-$1$ molecule on the surface of a three-dimensional topological insulator to be in the corresponding spin state as a function of $2\pi\hbar I/k_{\rm B}T$. We have set $2\pi\hbar D/k_{\rm B} T = 5$. The inset shows the maximum value of the probability $P_{1}$ for a molecule with $S=1$ as a function of $2 \pi \hbar D /k_B T$.}
    \label{fig:spin1}
\end{figure}

Remarkably, the complete polarization is lost again when the current is increased further. If $\jj \gg D (2 S - 1)$, not only electrons moving in the positive $z$ direction but essentially all electrons that contribute to the current $\vj$ can scatter off the localized magnetic moment and change its spin. For high current densities the transition rates are (for arbitrary direction of the anisotropy axis)
\begin{eqnarray}
  \Gamma_{s \pm 1,s} &\approx&
  \frac{J^2 \jj (S \mp s)(S \pm s+1)}{2 \pi} 
  F_{\pm}(\theta)
  \label{eq:GammaHigh3d}
\end{eqnarray}
where we abbreviated
\begin{eqnarray}
  F_{\pm}(\theta) &=& \frac{4(4 - \sin^2 \theta)}{3 \pi}
  \pm  \frac{\pi}{2} \cos \theta,
  \label{eq:Fdef}
\end{eqnarray}
and one quickly obtains the steady-state distribution from here,
\begin{equation}
  P_s = \frac{[F_+(\theta) - F_-(\theta)]
  F_+(\theta)^{S+s} F_-(\theta)^{S-s}}
  {F_+(\theta)^{2 S + 1} - F_-(\theta)^{2 S + 1}}.
  \label{eq:Ps3dHigh}
\end{equation}
Since $F_{\pm}(0) = (32 \pm 3 \pi^2)/6 \pi$, even for a perfectly aligned molecule no perfect polarization occurs in the limit of large currents. The probability $P_S$ to find the molecule in the maximal-spin state saturates at a value slightly above $96\%$ for molecules with large spin $S$. Although no perfect polarization results for current densities $\jj \gg D (2 S - 1)$, the rate at which the steady-state distribution is approached is considerably enhanced in comparison to the intermediate current-density regime, see Eq.\ (\ref{eq:GammaSwitch2dHigh}).

As an illustration, we show the probabilities $P_s$ for $s=-1,0,1$ for a molecule with $S=1$ as a function of the applied current density for $2\pi\hbar D/k_BT=5$, see Fig.\ \ref{fig:spin1}. Notice the (faint) maximum of $P_1$ for $\jj\sim D$. The inset shows that the maximum value of $P_1$ approaches unity in the limit when the anisotropy energy is much larger than the temperature, $\hbar D \gg k_{\rm B} T$.

\begin{figure}
\begin{center}
        \includegraphics[width=0.45\textwidth]{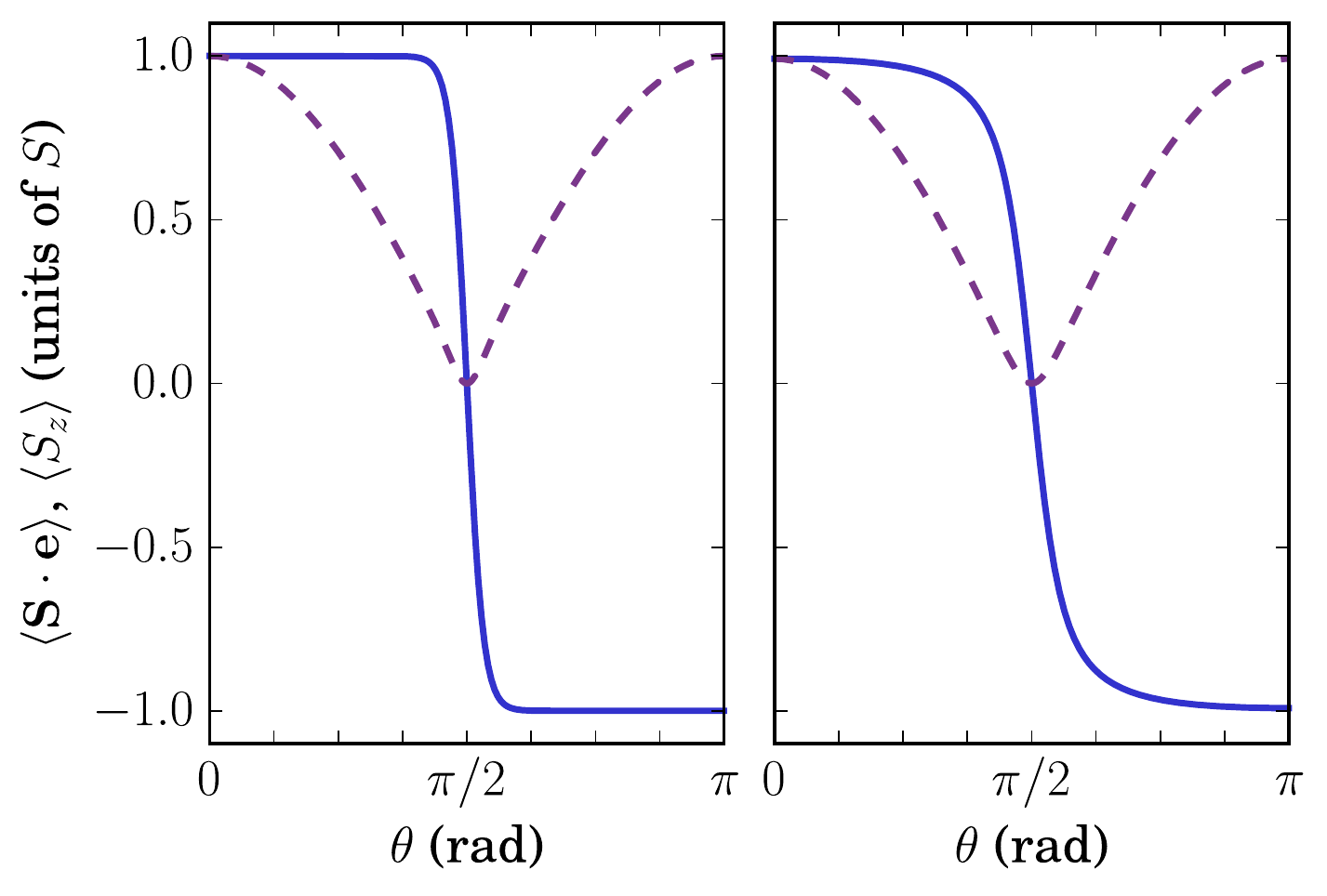}
    \end{center}
\caption{(Color online) Mean spin component $\langle \vS \cdot \ve \rangle$ along the direction of the anisotropy axis (solid curve) and the mean spin component $\langle S_z \rangle$ in the current direction (dashed), as a function of the angle $\theta$, for total spin $S=5$. Left panel: intermediate-current regime $k_{\rm B} T \ll \hbar \jj \ll \hbar D(2S-1)$; right panel: high-current regime $\jj \gg D(2S-1)$.}
  \label{fig:Mean3d}
\end{figure}

Figure \ref{fig:Mean3d} shows the mean spin components $\langle \vS \cdot \ve \rangle$ along the molecule's anisotropy axis and $\langle S_z \rangle$ along the current direction as a function of the angle $\theta$ for $S=5$, for the intermediate-current regime and for the high-current regime. As in the case of a two-dimensional TI, except for $\theta = \pi/2$, the application of a current always results in a net spin in the current direction. For a randomly oriented molecule, the average moment in the current direction approaches $S/2$ in the limit of large $S$.

\begin{figure}
\begin{center}
        \includegraphics[width=0.45\textwidth]{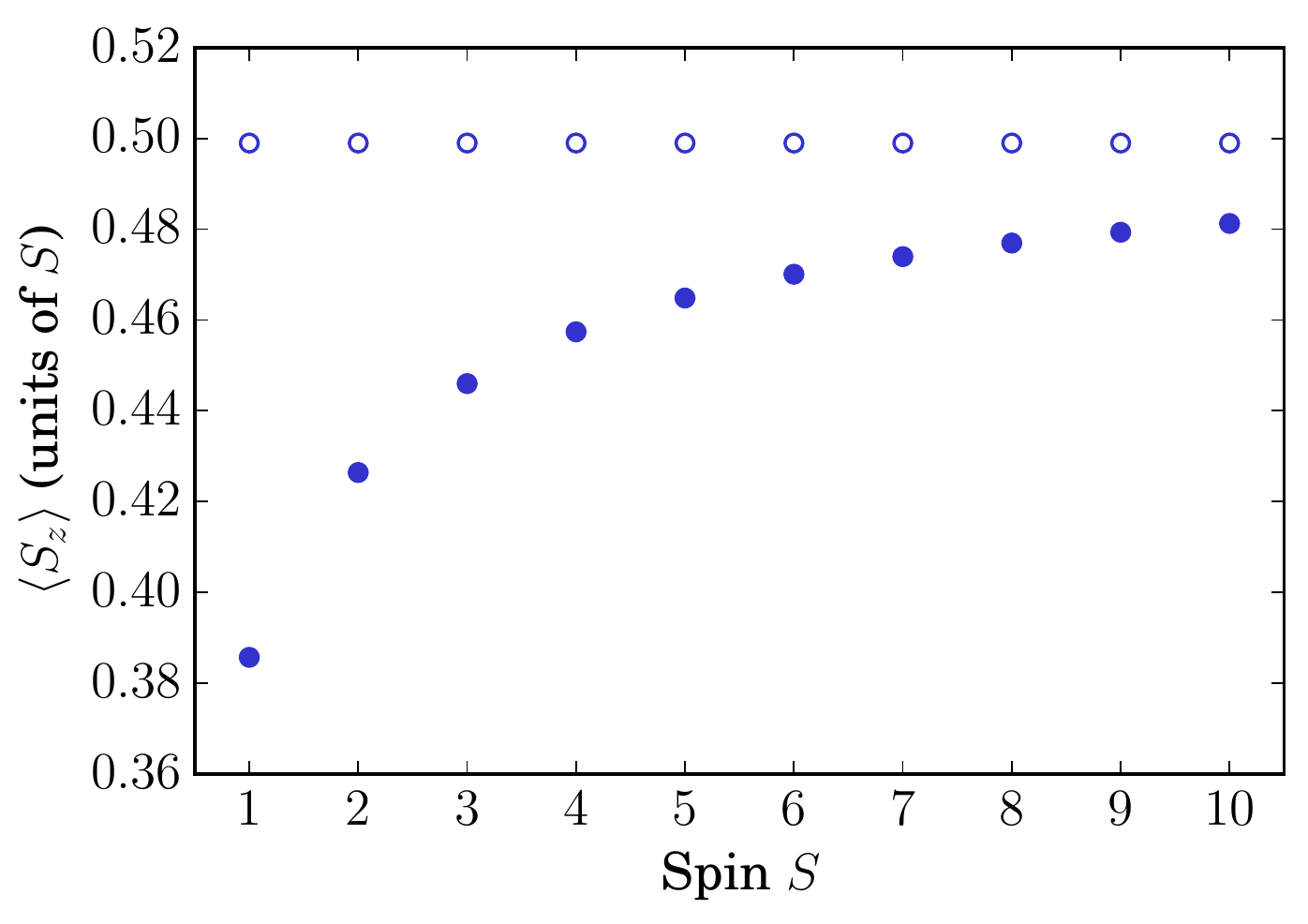}
    \end{center}
\caption{(Color online) Mean spin component in the current direction $\langle S_z \rangle$ per molecule as a function of the total spin $S$, for an ensemble of randomly oriented molecular magnets on the surface of a three-dimensional topological insulator. Open circles are for the intermediate-current regime $k_{\rm B} T \ll \hbar j \ll \hbar D(2S-1)$; filled circles are for the high-current regime $j \gg D(2S-1)$.}
  \label{fig:average}
\end{figure}

We can apply the results obtained here to a topological insulator surface with a dilute covering of molecular magnets. For a dilute covering our single-molecule analysis can be used to obtain the net polarization of the molecular layer. Assuming that the molecules have randomly orientated anisotropy axis --- a valid assumption if the magnetic core is shielded by an approximately spherical shell, see, {\em e.g.}, Ref.\ \onlinecite{hermanns_2013} --- the net polarization is found by averaging a single molecule's average moment over the directons of the anisotropy axis $\ve$. The calculation is straightforward in principle, starting from the steady-state distributions (\ref{eq:P3dEven}) and (\ref{eq:Ps3dHigh}). [See Eqs.\ (\ref{eq:P3dOdd}) and (\ref{eq:P3dHighOdd}) for steady-state distributions for half-integer spin $S$.] Instead of reporting the resulting expressions, which are rather cumbersome, we refer to Fig.\ \ref{fig:average}, which shows the resulting net polarization per molecule as a function of $S$.

\section{Conclusion}\label{sec:Conclusion}

The strict spin-momentum locking at the surface of a topological insulator provides an appealing mechanism to use electric currents to switch the spin state of a molecular magnet weakly coupled to the surface. Whereas based on simple spin conservation arguments one expects a complete response for a molecular magnet at the edge of a two-dimensional topological insulator if the molecule's magnetic anisotropy axis is aligned appropriately with the spin quantization axis of the surface state, the situation is more complicated for molecular magnets on the surface of a three-dimensional topological insulator and/or for an arbitrary orientation of the magnetic anisotropy axis. For a molecule on the surface of a three-dimensional topological insulator, a full current-induced polarization is achieved only if the molecule's magnetic anisotropy axis is aligned with the current direction and the magnitude $j$ of the current density is in an ``intermediate'' range $k_{\rm B} T \ll \hbar j/k_{\rm F} \ll \hbar D (2 S - 1)$, with $\hbar D S^2$ the molecule's anisotropy energy and $S$ the magnitude of its spin. Yet, even for a surface covered by molecules with randomly oriented anisotropy axes the application of a surface current should result in a net magnetic moment per molecule for all current strengths.
\\
\\
The minimal current density $j$ required to polarize the molecule and the resulting steady-state distribution of the molecule's magnetic states depends more on properties of the topological insulator surface than properties of the molecular magnet itself. 
The condition for an appreciable current-induced polarization is $j \gg k_{\rm B} T k_{\rm F}/\hbar$, {\em i.e.}, the disortion of the Fermi surface by the applied current should be large in comparison to the blurring of the Fermi surface by temperature. This is a condition that is easier to achieve if the chemical potential is in the vicinity of the Dirac point, so that $k_{\rm F}$ is small, and a relatively small current density implies a large shift of the distribution function. The role of the anisotropy energy is to set a crossover scale $j \sim D (2 S-1) k_{\rm F}$, above which current-induced transitions allow the molecule to be in all spin states, whereas for lower current densities only the magnetic ground states $\ket{S}$ and $\ket{-S}$ are accessible. Again, the crossover between these regimes happens at lower current densities if the chemical potential is closer to the Dirac point.
\\
\\
Details of the coupling between the molecule and the surface play a crucial role when it comes to setting the switching rate. This applies both to the zero-current spontaneous switching between the degenerate magnetic ground states $\ket{S}$ and $\ket{-S}$ and to the current-induced changes between these states. If the temperature is below the blocking temperature, appreciable switching rates can be obtained only in the ``high-current regime'' $j \gg D (2 S - 1) k_{\rm F}$, where we found the switching rate to be of order $\Gamma_{\rm switch} \sim J^2 j/k_{\rm F}$, where $J$ is the dimensionless exchange coupling. An upper limit for this rate is found by setting $J \sim 1$, although we expect that the presence of the organic shell will typically lead to much smaller exchange couplings for single-molecule magnets on a topological insulator surface. Taking a current density $j$ at the lower end of the high-current regime, $j \sim D (2 S-1) k_{\rm F}$, we find an upper bound for switching times in the range of $10^{-11}$ s for a molecule with a blocking temperature of a few K, but most likely significantly longer if $J$ is smaller. 
\\
\\
Our calculations have shown that in-plane currents can be used as an effective tool to switch the orientation of the spin of molecular magnets adsorbed on the surface of a three-dimensional topological insulator, even if no complete polarization is achieved for molecules without a specially aligned magnetic anisotropy axis. We hope that our observations, together with other theoretical studies highlighting the intriguing physics of molecular magnets on TI surfaces,\cite{owerre_2015} will motivate further experimental and theoretical research in this direction.

\subsection*{Acknowledgements}

We gratefully aknowledge support from the German Research Foundation (DFG) in the framework of the Collaborative Research Center 658.

\appendix

\section{Molecular magnets with half-integer spin}

In the case of a molecular magnet with half-integer spin $S$ the states $\ket{1/2}$ and $\ket{-1/2}$ are degenerate, which means that transitions between these states cannot be described by means of a rate equation, but that they need to be described with a master equation for the (reduced) density matrix 
\begin{equation}
  \rho = \left( 
  \begin{array}{cc} \rho_{1/2,1/2} & \rho_{1/2,-1/2} \\ 
  \rho_{1/2,-1/2} & \rho_{-1/2,-1/2} \end{array} \right).
  \label{eq:rhoDef}
\end{equation}
The diagonal elements $\rho_{s,s} = P_s$ are the probabilities to find the molecule in spin state $\ket{s}$, $s = \pm 1/2$; the off-diagonal elements describe coherences beween these two spin states. 

To allow for a unified treatment of a molecular magnet coupled to the edge of a two-dimensional TI and a molecular magnet at the surface of a three-dimensional TI, we write matrix elements of the exchange Hamiltonian (\ref{eq:Hex2d}) for the two-dimensional case as
\begin{equation}
  \bra{\varphi',s'} H_{\rm  ex} \ket{\varphi,s} =
  \frac{1}{2} J v \hbar m_{s',s}(\varphi',\varphi),
\end{equation}
where we use the propagation angle $\varphi = 0$ for a right-moving electron (labeled with $\tau = +1$ in Sec.\ \ref{sec:2dTI}) and $\varphi = \pi$ for a left-moving electron (labeled with $\tau = -1$). Matrix elements of the exchange Hamiltonian (\ref{eq:Hex3d}) for the three-dimensional case are written
\begin{equation}
  \bra{\varphi',s'} H_{\rm  ex} \ket{\varphi,s} =
  \frac{J v \hbar}{k_{\rm F}} m_{s',s}(\varphi',\varphi).
\end{equation}
The rates $\Gamma_{s',s}$, see Eqs.\ (\ref{eq:GammaAnis}) and (\ref{eq:GammaGeneral}), are then expressed as
\begin{widetext}
\begin{eqnarray}
  \Gamma_{s',s} &=&
  J^2 \int \frac{d\varphi}{2 \pi} \frac{d\varphi'}{2 \pi}
  |m_{s',s}(\varphi',\varphi)|^2 g[\jj (\cos \varphi'-\cos\varphi)/2 + D(s^2 - s'^2)], \label{eq:GammassApp}
\end{eqnarray}
where the double integration $\int (d\varphi/2\pi) (d\varphi'/2 \pi)$ should be replaced by a double sum $(1/4) \sum_{\varphi,\varphi' = 0, \pi}$ for the case of a two-dimensional TI. The explicit expressions for the $m(\varphi',\varphi)$ read
\begin{eqnarray}
  \hbar m(\varphi',\varphi) &=&
  2 (S_z \cos \theta - S_x \sin \theta) \cos\varphi_+ +
  (S_x \cos \theta + S_z \sin \theta)
    (\cos \phi \sin \varphi_+ + i \sin \phi \sin \varphi_-)
  \nonumber \\ && \mbox{}
  + i S_y (\cos \phi \sin \varphi_- + i \sin \phi \sin \varphi_+),
\end{eqnarray}
where $\varphi_{\pm} = (\varphi' \pm \varphi)/2$, see the text below Eq.\ (\ref{eq:MatrixElements3d}). 

Because of the large energy differences between spin states with different values of $|s|$ and the lack of direct transitions between the degenerate spin states $\ket{s}$ and $\ket{-s}$ for $|s| > 1/2$, the rates (\ref{eq:GammassApp}) are sufficient to calculate the (rate of change of the) probabilities $P_s$ that the molecule is in spin state $\ket{s}$ for $|s| > 1/2$, see Eq.\ (\ref{eq:master_time-dep}). On the other hand, for the degenerate spin states $\ket{1/2}$ and $\ket{-1/2}$ we must use the full density matrix $\rho$, see Eq.\ (\ref{eq:rhoDef}). The equation of motion for this density matrix has the well-known Lindblad form\cite{kossakowski,lindblad_1976}
\begin{eqnarray}
  \frac{d\rho}{dt} &=&
  - \frac{i}{\hbar} 
  (\overline {H_{\rm ex}} \rho - \rho \overline {H_{\rm ex}}) 
  + \sum_{\pm} \left[ \Gamma_{\pm 1/2,\pm 3/2} \Pi_{\pm 1/2} P_{\pm 3/2}
  - \frac{1}{2} \Gamma_{\pm 3/2,\pm 1/2} \left(
  \Pi_{\pm 1/2} \rho + \rho \Pi_{\pm 1/2} \right) \right]
  \label{eq:Lindblad}
  \\ && \mbox{}
  + J^2 \int \frac{d\varphi}{2 \pi} \frac{d\varphi'}{2 \pi}
  g[\jj (\cos \varphi'-\cos\varphi)/2]
  \left\{ m(\varphi',\varphi) \rho m(\varphi',\varphi)^{\dagger}
  - \frac{1}{2} \left[
  m(\varphi',\varphi)^{\dagger} m(\varphi',\varphi) \rho
  +
  \rho m(\varphi',\varphi)^{\dagger} m(\varphi',\varphi) \right]
  \right\}, \nonumber
\end{eqnarray}
\end{widetext}
where the projection matrices $\Pi_{1/2}$, and $\Pi_{-1/2}$ are defined as
\begin{equation}
  \Pi_{1/2} = \left( 
  \begin{array}{cc} 1 & 0 \\ 0 & 0 \end{array} \right),\ \
  \Pi_{-1/2} = \left( 
  \begin{array}{cc} 0 & 0 \\ 0 & 1 \end{array} \right).
\end{equation}
One verifies that the Lindblad equation (\ref{eq:Lindblad}) simplifies to the standard rate equation (\ref{eq:master_time-dep}) if the density matrix $\rho$ is a diagonal matrix, $\rho = \mbox{diag}\,(P_{1/2},P_{-1/2})$.

The first term in Eq.\ (\ref{eq:Lindblad}) contains the expectation value of the exchange Hamiltonian projected onto the spin states $\ket{\pm 1/2}$. It acts as an effective magnetic field driving coherent transitions between the states $\ket{1/2}$ and $\ket{-1/2}$. This term plays no role for states $\ket{s}$ with $|s| > 1/2$ because it is much smaller than the anisotropy field if the exchange coupling is weak, $|J| \ll 1$. However, for transitions between the states $\ket{1/2}$ and $\ket{-1/2}$ there is no anisotropy field, and the exchange field competes with the similarly small scattering-induced transitions between the states $\ket{1/2}$ and $\ket{-1/2}$. With the same notation as above we have
\begin{eqnarray}
  \overline{H_{\rm ex}} &=&
  \frac{J \jj \hbar}{2} \int \frac{d\varphi}{2 \pi} m(\varphi,\varphi) \cos \phi,
\end{eqnarray}
where, for a molecule coupled to the edge of a two-dimensional topological insulator, the integration $\int (d\varphi/2\pi)$ should be replaced by a summation $(1/2) \sum_{\varphi = 0,\pi}$. Performing the angular integration gives
\begin{equation}
  \overline{H_{\rm ex}} =
  \frac{J \jj \hbar}{8}
  \left( \begin{array}{cc}
  2 \cos \theta & - (2 S + 1) \sin \theta \\ 
  - (2 S+1) \sin \theta & - 2 \cos \theta
  \end{array} \right),
  \label{eq:HexExpr}
\end{equation}
which is indeed what one expects for an exchange field in the $z$ direction, recalling that the spin states $\ket{s}$ are defined with respect to the anisotropy axis $\ve = \cos \theta \ve_z + \sin \theta \cos \phi \ve_x + \sin \theta \sin \phi \ve_y$.

A solution of the Lindblad equation proceeds by parameterizing 
\begin{equation}
  \rho = \frac{1}{2} \rho_0 \tau_0 + \frac{1}{2} \sum_{j=x,y,z} \rho_j \tau_j, 
\end{equation}
where $\tau_0$ is the $2 \times 2$ identity matrix and the $\tau_j$ are the Pauli matrices in the space spanned by $\ket{1/2}$ and $\ket{-1/2}$. We give the equation of motion for $\rho$ for the intermediate and high current regimes, in which we may approximate $g[\jj (\cos \varphi'-\cos\varphi)/2] \approx \jj (\cos \varphi-\cos\varphi')/2$ if $\cos \varphi > \cos \varphi'$ and $g[\jj (\cos \varphi'-\cos\varphi)/2] \approx 0$ otherwise (including the case $\cos \varphi = \cos \varphi'$). 

In the two-dimensional case there is a contribution from $\varphi = 0$, $\varphi = \pi$ only. The equations of motion read
\begin{widetext}
\begin{eqnarray}
  \frac{d \rho_0}{dt} &=&
  \sum_{\pm} \left[ \Gamma_{\pm 1/2,\pm 3/2} P_{\pm 3/2}
  - \frac{1}{2} \Gamma_{\pm 3/2,\pm 1/2} (\rho_0 \pm \rho_z) \right],
  \nonumber \\
  \frac{d\rho_x}{dt} &=& - \frac{J \jj}{8} \rho_y \cos \theta 
  + \frac{J^2 \jj}{8} \left\{-\rho_x [4 \sin^2 \theta + (2 S+1)^2]
  + 2 (\rho_z \cos \theta - 2 \rho_0) (2 S+1) \sin \theta \right\} 
  - \frac{1}{2} \rho_x \sum_{\pm} \Gamma_{\pm 3/2,\pm 1/2}, \nonumber \\
  \frac{d \rho_y}{dt} &=& 
  \frac{J \jj}{16}\left[ 2 \rho_x \cos \theta + \rho_z (2 S+1) \sin \theta
  \right] 
  - \frac{J^2 \jj}{8} \rho_y
  \left[ 4 \sin^2 \theta + (2 S+1)^2 \cos^2 \theta \right]
  - \frac{1}{2} \rho_y \sum_{\pm} \Gamma_{\pm 3/2,\pm 1/2}, \nonumber \\
  \frac{d \rho_z}{dt} &=& 
  - \frac{J \jj}{16} (2 S+1) \rho_y \sin \theta
  + \frac{J^2 \jj}{8} (2 S+1)
  \left\{ (2 S+1)[2 \rho_0 \cos \theta - \rho_z (\cos^2 \theta + 1)]
  + 2 \rho_x \sin \theta \cos \theta \right\} \nonumber \\ && \mbox{}
  + \sum_{\pm} \left[ \pm \Gamma_{\pm 1/2,\pm 3/2} P_{\pm 3/2}
  - \frac{1}{2} \Gamma_{\pm 3/2,\pm 1/2} (\rho_z \pm \rho_0)\right], 
  \label{eq:Lindblad2d}
\end{eqnarray}
In the three-dimensional case the equations of motion are
\begin{eqnarray}
  \frac{d \rho_0}{dt} &=&
  \sum_{\pm} \left[ \Gamma_{\pm 1/2,\pm 3/2} P_{\pm 3/2}
  - \frac{1}{2} \Gamma_{\pm 3/2,\pm 1/2} (\rho_0 \pm \rho_z) \right],
  \nonumber \\
  \frac{d\rho_x}{dt} &=& - \frac{J \jj}{8} \rho_y \cos \theta
  - \frac{J^2 \jj}{3 \pi^2}
   \left\{ 
  \rho_x \left[ 4 + \sin^2 \theta + 2 (2 S+1)^2 \right] 
  + 4 \sin \theta \cos \theta \rho_z \right\}
  - \frac{J^2 \jj}{4} \rho_0 (2 S+1) \sin \theta 
  \nonumber \\ && \mbox{}
  - \frac{1}{2} \rho_x \sum_{\pm} \Gamma_{\pm 3/2,\pm 1/2}, \nonumber \\ 
  \frac{d \rho_y}{dt} &=&
  \frac{J \jj}{16}
  \left[ 2 \rho_x \cos \theta + \rho_z (2 S+1) \sin \theta \right] 
  - \frac{J^2 \jj}{3 \pi^2} \rho_y
  \left[4 + 4 \sin^2 \theta + (2 S+1)^2(1 + \cos^2 \theta) \right]
  \nonumber \\ && \mbox{}
  - \frac{1}{2} \rho_y \sum_{\pm} \Gamma_{\pm 3/2,\pm 1/2}, \nonumber \\
  \frac{d \rho_z}{dt} &=& 
  - \frac{J \jj}{16} (2 S+1) \rho_y \sin \theta
  + \frac{J^2 \jj}{3 \pi^2} (2S+1)
  \left[2 \rho_x \sin \theta \cos \theta - \rho_z (2 S+1) (4 - \sin^2 \theta) \right]
  + \frac{J^2 \jj}{8} (2S+1)^2 \rho_0 \cos \theta
  \nonumber \\ && \mbox{}
  + \sum_{\pm} \left[ \pm \Gamma_{\pm 1/2,\pm 3/2} P_{\pm 3/2}
  - \frac{1}{2} \Gamma_{\pm 3/2,\pm 1/2} (\rho_z \pm \rho_0)\right].
  \label{eq:Lindblad3d}
\end{eqnarray}
\end{widetext}
In the intermediate-current regime we may set 
\begin{equation}
  \Gamma_{\pm 3/2,\pm 1/2} \approx \frac{J^2 D}{2} (2 S-1)(2 S+3),
\end{equation}
see Eqs.\ (\ref{eq:GammaAnis}) and (\ref{eq:Gamma3dIntermediate}). For the high-current regime we may approximate $\Gamma_{\pm 3/2,\pm 1/2}$ by Eqs.\ (\ref{eq:GammaHigh2d}) and (\ref{eq:GammaHigh3d}) for the two-dimensional and three-dimensional cases, respectively.

When searching for a stationary solution, the contributions from transitions to/from the spin states $\ket{\pm 3/2}$ vanish in the equations of motion for $\rho_0$ and $\rho_z$ because of detailed balance. They remain, however, in the equations of motion for $\rho_x$ and $\rho_y$. Once the ratio $P_{1/2}/P_{-1/2}$ has been determined from the stationary solution of the Lindblad equations (\ref{eq:Lindblad2d}) or (\ref{eq:Lindblad3d}), the remaining probabilities $P_s$ for $|s| > 1/2$ can be found from detailed balance, 
\begin{equation}
  P_{|s|} = \frac{\Gamma_{|s|,|s|-1} P_{|s|-1}}{\Gamma_{|s|-1,|s|}},\ \
  s = \pm 3/2,\ldots,\pm S,
\end{equation}
and from the normalization condition $\sum_{s} P_s = 1$.

We now report expressions for the ratio $P_{1/2}/P_{-1/2}$ for the intermediate and high current regimes $k_{\rm B} T \ll \hbar \jj \ll \hbar D(2 S - 1)$ and $\jj \gg D (2 S - 1)$ for a spin coupled to the edge of a two-dimensinal TI and to the surface of a three-dimensional TI. The expressions for the intermediate-current regime are simplified using the inequalities $\jj \ll D (2 S - 1)$ and $J \ll 1$, but we make no assumptions regarding the relative magnitude of these two quantities. (Note that the intermediate current regime does not exist for $S=1/2$.) The expressions for the high-current limit are simplified using the inequality $J \ll 1$.

In the intermediate-current regime we find for the case of a molecular magnet coupled to the edge of a two-dimensional TI
\begin{equation}
  \frac{P_{1/2}}{P_{-1/2}} = \frac{G_+(\theta)}{G_-(\theta)},
\end{equation}
with
\begin{equation}
  G_{\pm}(\theta) =
  (1 \pm \cos \theta)^2 + \frac{\jj \sin^2 \theta}{16 D J^2(2S-1)(2S+3)}.
  \label{eq:Gplus2dIntermed}
\end{equation}
This result agrees with what one obtains from naive application of the rate equations if $\jj \ll J^2 D (2 S-1)$, but differs otherwise because of the effect of the current-induced exchange field (\ref{eq:HexExpr}). Note that $G_-(\theta) = G_+(\pi - \theta)$. The resulting steady-state probability distribution is
\begin{eqnarray}
   P_{S} &=& 1 - P_{-S} \\ &=& \nonumber 
   \frac{G_+(\theta)(1 + \cos \theta)^{4 S-2}}{G_+(\theta)(1 + \cos \theta)^{4 S-2}
  + G_-(\theta)(1 - \cos \theta)^{4 S-2}}.
  \label{eq:P2dIntermediateOdd}
\end{eqnarray}
The current-induced switching rate from $\ket{-S}$ to $\ket{S}$ can be calculated as the transition rate from $\ket{-1/2}$ to $\ket{1/2}$ starting with the initial condition 
\begin{eqnarray*}
  \rho(0) &=& \frac{(1 + \cos \theta)^{S-1/2}}{2^{S-1/2}} \Pi_{-1/2} \nonumber \\ && \mbox{} \times e^{2 \pi \hbar (S-1/2)(\jj - D (S+1/2))/k_{\rm B} T},
\end{eqnarray*}%
which is the reduced density matrix after current-induced equilibration of all states with $s < 0$, including the state $\ket{-1/2}$, but without including transitions from $\ket{-1/2}$ to $\ket{1/2}$. Within a  time $\sim 2/J^2 D (2 S-1)(2 S + 3)$, which is short in comparison to the time required to transition from to $\ket{-1/2}$ to $\ket{1/2}$, a quasi-steady state is reached for the reduced density matrix $\rho$, in which the off-diagonal component $\rho_y$ becomes nonzero, 
\begin{eqnarray*}
  \rho_y \approx \frac{\rho_z \jj (2 S+1)}{8 J D (2 S - 1)(2 S + 3)}.
\end{eqnarray*}
The current-induced switching rate $\Gamma_{\rm switch}(\jj)$ is then found as the rate of change of $P_{1/2}$ in this quasi-steady state, which gives
\begin{eqnarray}
  \Gamma_{\rm switch}(\jj) &=& 
  \frac{J^2 \jj (2 S+1)^2}{2^{S+7/2}}
  G_+(\theta) (1 + \cos \theta)^{S-1/2} 
  \nonumber \\ && \mbox{} \times
  e^{2 \pi \hbar (S-1/2)(\jj - D (S+1/2))/k_{\rm B} T},
\end{eqnarray}
where $G_+(\theta)$ was defined in Eq.\ (\ref{eq:Gplus2dIntermed}). The switching rate for the opposite process is given the same expression, but with the replacement $\theta \to \pi - \theta$.
  
For the high-current limit and a two-dimensional TI we find $P_{1/2}/P_{-1/2} = G_+(\theta)/G_-(\theta)$ with the function $G_{\pm}(\theta)$ defined as
\begin{equation}
  G_{\pm}(\theta) = 
  \frac{1 \pm \cos \theta}{2} + \frac{3(2 S-1)(2 S + 3)}{32} \sin^2 \theta,
\end{equation}
which is different from the result one obtains from naive application of the rate equations. For $S=1/2$ the ratio $P_{1/2}/P_{-1/2}$ is consistent with a spin polarized in the $z$ direction. The resulting probability distribution $P_s$ reads
\begin{eqnarray}
  P_{\pm|s|} &=& \frac{4 \cos \theta G_{\pm}(\theta)} 
  {G_+(\theta) (1 + \cos \theta)^{2 S-1}
    + G_-(\theta) (1 - \cos \theta)^{2 S-1}}
  \nonumber \\ && \mbox{} \times
  \frac{(1 \pm \cos \theta)^{2(|s|+S-1)} (1 \mp \cos \theta)^{2(S-|s|)}}
  {(1 + \cos\theta)^{2 S+1} - (1 - \cos \theta)^{2 S+1}}.
\end{eqnarray}
Equation (\ref{eq:GammaSwitch2dHigh}) of the main text is a good order-of-magnitude estimate for the rate $\Gamma_{\rm switch}(\jj)$ at which this steady-state distribution is approached.

For a molecule on the surface of a three-dimensional TI with an applied current in the intermediate-current regime $k_{\rm B} T \ll \hbar \jj \ll \hbar D (2 S -1)$ the stationary solution of the Lindblad equation (\ref{eq:Lindblad3d}) gives the ratio $P_{1/2}/P_{-1/2} = G_+(\theta)/G_-(\theta)$ with
\begin{eqnarray}
  G_{\pm}(\theta) &=& F_{\pm}(\theta)
  + \frac{\pi \jj \sin^2 \theta}{32 D J^2 (2 S - 1)(2 S + 3)},
\end{eqnarray}
where the function $F_{\pm}(\theta)$ was defined in Eq.\ (\ref{eq:Fdef}) of the main text. This expression simplifies to the ratio $P_{1/2}/P_{-1/2} = F_+(\theta)/F_-(\theta)$ one obtains by considering rates only in the limit $\jj \ll D J^2 (2 S - 1)$. The corresonding steady-state distribution reads
\begin{eqnarray}
   P_{S} &=& 1 - P_{-S} \nonumber \\ &=&
   \frac{G_+(\theta) F_+(\theta)^{4 S-2}}{G_+(\theta)F_+(\theta)^{4 S-2}
  + G_-(\theta)F_-(\theta)^{4 S-2}}.
  \label{eq:P3dOdd}
\end{eqnarray}
The switching rate from $\ket{-S}$ to $\ket{S}$ is calculated in the same way as for the case of a two-dimensional TI, and one finds
\begin{eqnarray}
 \Gamma_{\rm switch}(\jj) &=&
  \frac{J^2 \jj (2 S+1)^2 G_+(\theta)}{8 \pi}
  \nonumber \\ && \mbox{} \times
  \left( \frac{2 (1 + \cos \theta)^2 
  k_{\rm B} T}{\pi^2 \hbar \jj} \right)^{S-1/2}
  \nonumber \\ && \mbox{} \times
  e^{2 \pi \hbar(S-1/2) (\jj - D (S + 1/2))/k_{\rm B} T}.
  \label{eq:GammaSwitchJ3dOdd}
\end{eqnarray}
The switching rate for the inverse transition is again given by the replacement $\theta \to \pi - \theta$.

Finally, for the high-current regime in three dimensions we find that the stationary solution of the Lindblad equation (\ref{eq:Lindblad3d}) has $P_{1/2}/P_{-1/2} = G_+(\theta)/G_-(\theta)$ with
\begin{eqnarray}
  G_{\pm}(\theta) &=& \frac{1}{2} \pm \frac{3 \pi^2}{64} \cos \theta
  \\ && \nonumber \mbox{}
  + \frac{(2 S - 1)(2 S + 3)}{64} (8 - \sin^2 \theta) \sin^2 \theta.
\end{eqnarray}
The corresponding steady-state distribution reads
\begin{eqnarray}
  P_{\pm|s|} &=& \frac{F_+(\theta) - F_-(\theta)}
  {G_+(\theta) F_+(\theta)^{S-1/2}
    + G_-(\theta) F_-(\theta)^{S-1/2}}
  \nonumber \\ && \mbox{} \times
  \frac{G_{\pm}(\theta) F_{\pm}(\theta)^{|s|+S-1} F_{\mp}(\theta)^{S-|s|}}{
  F_+(\theta)^{S+1/2} - F_-(\theta)^{S+1/2}}.
  \label{eq:P3dHighOdd}
\end{eqnarray}
This distribution is approached at the rate given by Eq.\ (\ref{eq:GammaSwitch2dHigh}) of the main text.

If the magnetic anisotropy axis is aligned with the current direction, the rate equation approach can be used throughout. In this case the transition rates $\Gamma_{1/2,-1/2}$ and $\Gamma_{-1/2,1/2}$ for the intermediate and high-current regimes read
\begin{eqnarray}
  \Gamma_{\pm 1/2, \mp 1/2} &=& 
  J^2 \jj (S+1/2)^2 
  \left[ \frac{8}{3 \pi^2} \pm \frac{1}{4} \right].
\end{eqnarray}
In the intermediate-current regime, the steady-state spin polarization of the molecule is complete; The probability $P_{-S}$ vanishes $\propto (k_{\rm B} T/\hbar \jj)^{2 S - 1}$ if $S$ is half integer, whereas all other probabilities are exponentially small in $\hbar D (2 S - 1)/k_{\rm B} T$. The switching rate out of the state $\ket{-S}$ is, in the intermediate-current regime,
\begin{eqnarray}
 \Gamma_{\rm switch}(\jj) &=&
  \frac{J^2 \hbar \jj (S+1/2)^2(32 + 3 \pi^2)}{12 \pi^2 \hbar}
  \left( \frac{8 k_{\rm B} T}{\pi^2 \hbar \jj} \right)^{S-1/2}
  \nonumber \\ && \mbox{} \times
  e^{2 \pi \hbar(S-1/2) (\jj - D (S + 1/2)/k_{\rm B} T}.
  \label{eq:GammaSwitchJ3dOddAligned}
\end{eqnarray}
In the high-current regime the order-of-magnitude estimate (\ref{eq:GammaSwitch2dHigh}) for the switching rate also applies to the half-integer spin case.

\end{document}